\documentclass[preprint,12pt]{elsarticle}
\usepackage{amssymb}
\journal{Nuclear Physics A}

\begin{document}

\begin{frontmatter}



\title{Time-dependent
Ginzburg-Landau Equation in the Nambu--Jona-Lasinio Model}


\author[label1]{Wei-jie Fu}
\author[label1]{Da Huang}
\author[label1]{Fa-bing Wang}

\address[label1]{Kavli Institute for
Theoretical Physics China (KITPC), Key Laboratory of Frontiers in
Theoretical Physics, Institute of Theoretical Physics, Chinese
Academy of Science, Beijing 100190, China}

%\email[]{wjfu@itp.ac.cn}

\begin{abstract}
We apply the closed time-path Green function formalism in the
Nambu--Jona-Lasinio model. First of all, we use this formalism to
obtain the well-known gap equation for the quark condensate in a
stationary homogeneous system. We have also used this formalism to
obtain the Ginzburg-Landau (GL) equation and the time-dependent
Ginzburg-Landau (TDGL) equation for the chiral order parameter in an
inhomogeneous system. In our derived GL and TDGL equations, there is
no other parameters except for those in the original NJL model.
\end{abstract}


\end{frontmatter}

% \linenumbers

\section{Introduction}
\vspace{5pt}

QCD thermodynamics, for example the equation of state of quark gluon
plasma (QGP), phase transition of chiral symmetry restoration,
deconfinement phase transition and so on, has been a subject of
intensive investigations in recent years. On the one hand, the
deconfined QGP are expected to be formed in ultrarelativistic
heavy-ion
collisions~\cite{Shuryak2004,Gyulassy2005,Shuryak2005,Arsene2005,Back2005,Adams2005,Adcox2005,Blaizot2007}
(for example the experiments at the Relativistic Heavy Ion Collider
(RHIC) and at the Large Hadron Collider (LHC)) and in the interior
of neutron stars~\cite{Weber2005,Alford2007,Alford2008,Fu2008b}; On
the other hand, studying the thermodynamical behaviors of the QGP,
especially the deconfinement and chiral phase transitions, is an
elementary problem in strong interaction physics.

Lattice QCD
calculations~\cite{Fodor2002,Ejiri2004,Gavai2005,deForcrand2003} as
well as models
studies~\cite{Asakawa1989,Barducci1989,Barducci1994,Berges1999,Halasz1998,Scavenius2001,Hatta2003,Barducci2005,Fu2008}
indicate that there is a critical point in the QCD phase diagram in
the plane of temperature and baryon chemical
potential~\cite{Stephanov2006}, which separates the first order
phase transition at high baryon chemical potential from the
continuous crossover at high temperature. Furthermore, experiments
with the goal to search for the QCD critical point are planed and
underway at RHIC  and at the Super Proton Synchrotron (SPS)
~\cite{Mohanty2009,Anticic2009,Aggarwal2010,Aggarwal2010b}.
Therefore, the critical dynamics of the chiral phase transition has
attracted lots of attentions in recent years. Based on equilibrium
thermodynamics~\cite{Stephanov1998} or quasi-stationary
framework~\cite{Stephanov2010}, signatures of the QCD critical point
have been studied by M. A. Stephanov~\textit{et al.}. A.
Singh~\textit{et al.} have also studied the kinetics of the chiral
phase transition subsequent to a quench from a disordered phase to
the ordered phase~\cite{Singh2010}.

In this work, we will adopt the closed time-path Green function
(CTPGF) formalism to derive the Ginzburg-Landau equation and the
time-dependent Ginzburg-Landau equation of the order parameter for
the chiral phase transition in an inhomogeneous quark matter system
in the Nambu--Jona-Lasinio (NJL) model. The CTPGF formalism,
developed by Schwinger~\cite{Schwinger1961} and
Keldysh~\cite{Keldysh1965}, has been used to solve lots of
interesting problems in statistical physics and condensed matter
theory~\cite{Chou1985}, and it has also been used in the NJL model
to derive the transport equations~\cite{Klevansky1997}. It is
generally believed that this technique is quite effective in
investigating the nonequilibrium statistical
theory~\cite{Chou1985,Zhou1980}. It has also been used to treat a
system of self-interacting bosons described by $\lambda\phi^{4}$
scalar fields~\cite{Calzetta1988}.

The paper is organized as follows. In Sec.~\ref{Sec2} we simply
review the CTPGF formalism, mainly on the generating functional of
the CTPGFs. In Sec.~\ref{Sec3} we apply the CTPGF formalism into the
NJL model. First of all, we use this formalism to derive the
well-known gap equation of the quark condensate for the homogeneous
system, then we also obtain the Ginzburg-Landau equation and
time-dependent Ginzburg-Landau equation of the order parameter for
an inhomogeneous system. In Sec.~\ref{Sec4} we present our summary
and conclusions.

\section{Simple Review on the CTPGF}
\label{Sec2}

\subsection{Two-point CTPGFs}

In this section we give a short review about the CTPGF formalism and
introduce some notations which we will use in our following
discussions. Readers who are not familiar with the CTPGF formalism
are strongly suggested to reference the excellent review article by
K.~C.~Chou~\textit{et al.}~\cite{Chou1985}. Here, we take the real
boson field $\varphi (x)$ for example. The two-point CTPGF is
defined as
\begin{eqnarray}
G(x,y)&\equiv&-i\mathrm{Tr}\{T_{p}(\varphi(x)\varphi(y))\hat{\rho}\}\nonumber\\
&\equiv&-i\langle T_{p}(\varphi(x)\varphi(y))\rangle,\label{Gp}
\end{eqnarray}
where $\hat{\rho}$ is the density matrix and $T_{p}$ is the time
ordering operator along the closed time-path $p$ which goes from
$-\infty$ to $+\infty$ (also denoted as the positive time branch)
and then returns back from $+\infty$ to $-\infty$ (negative time
branch). We should note that any event at the negative time branch
is later than any that at the positive time branch. The Green
function $G(x,y)$ in Eq.(\ref{Gp}) on the closed time-path $p$ can
be expressed as Green function whose coordinates $x$ and $y$ are
constrained to be on either positive or negative time branches, and
then $G(x,y)$ becomes a $2\times 2$ matrix, i.e.,
\begin{equation}
\hat{G}(x,y)\equiv\Bigg(\begin{array}{cc} G_{++} & G_{+-}\\
G_{-+} & G_{--}\end{array}\Bigg)\equiv\Bigg(\begin{array}{cc} G_{F} & G_{+}\\
G_{-} & G_{\tilde{F}}\end{array}\Bigg)\label{hatG}
\end{equation}
with
\begin{eqnarray}
G_{F}(x,y)&\equiv&-i\langle T(\varphi(x)\varphi(y))\rangle,  \\
G_{+}(x,y)&\equiv&-i\langle\varphi(y)\varphi(x)\rangle,\\
G_{-}(x,y)&\equiv&-i\langle\varphi(x)\varphi(y)\rangle,\\
G_{\tilde{F}}(x,y)&\equiv&-i\langle
\tilde{T}(\varphi(x)\varphi(y))\rangle,
\end{eqnarray}
where $T$ is the usual time-ordering operator and $\tilde{T}$ is
anti-time-ordering operator. Therefore, $G_{F}$ is the conventional
Feynman causal propagator and $G_{\tilde{F}}$ is an anti-causal
propagator, whose expression can be given explicitly as
\begin{equation}
G_{\tilde{F}}(x,y)=-i\theta(y_{0},
x_{0})\langle\varphi(x)\varphi(y)\rangle-i\theta(x_{0},
y_{0})\langle\varphi(y)\varphi(x)\rangle.
\end{equation}
Furthermore, It can be easily proved that we have the following
identity, i.e.,
\begin{equation}
G_{F}(x,y)+G_{\tilde{F}}(x,y)=G_{+}(x,y)+G_{-}(x,y).
\end{equation}
So far we have presented CTPGFs in two representations: one is the
closed time-path representation, i.e., Eq.(\ref{Gp}) and the other
is the single time representation given by Eq.(\ref{hatG}). In fact,
there is another representation which is directly related with
measurable quantities, which is often called as physical
representation and defined as
\begin{eqnarray}
G_{r}(x,y)&\equiv&-i\theta(x_{0},
y_{0})\langle [\varphi(x),\varphi(y)]\rangle,  \\
G_{a}(x,y)&\equiv& i\theta(y_{0},
x_{0})\langle [\varphi(x),\varphi(y)]\rangle,\\
G_{c}(x,y)&\equiv&-i\langle \{\varphi(x),\varphi(y)\}\rangle.
\end{eqnarray}
$G_{r}$, $G_{a}$, and $G_{c}$ are retarded, advanced, and
correlation Green functions, respectively. The relations between the
CTPGFs in the physical representation and those in the single time
representation are given by
\begin{eqnarray}
G_{r}&=&G_{F}-G_{+}=G_{-}-G_{\tilde{F}},\\
G_{a}&=&G_{F}-G_{-}=G_{+}-G_{\tilde{F}},\\
G_{c}&=&G_{F}+G_{\tilde{F}}=G_{+}+G_{-},
\end{eqnarray}
and the inverse relations are
\begin{equation}
\hat{G}=\frac{1}{2}G_{r}\Bigg(\begin{array}{cc} 1 & -1\\
1 & -1\end{array}\Bigg)+\frac{1}{2}G_{a}\Bigg(\begin{array}{cc} 1 & 1\\
-1 & -1\end{array}\Bigg)+\frac{1}{2}G_{c}\Bigg(\begin{array}{cc} 1 & 1\\
1 & 1\end{array}\Bigg).\label{hatG2}
\end{equation}
Introducing two-component vectors
\begin{equation}
\xi\equiv\Bigg(\begin{array}{c} 1 \\
1 \end{array}\Bigg),\qquad\eta\equiv\Bigg(\begin{array}{c} 1 \\
-1 \end{array}\Bigg),\label{}
\end{equation}
ones can express Eq.(\ref{hatG2}) as
\begin{equation}
G_{\alpha\beta}=\frac{1}{2}G_{r}\xi_{\alpha}\eta_{\beta}+\frac{1}{2}G_{a}\eta_{\alpha}\xi_{\beta}+\frac{1}{2}G_{c}\xi_{\alpha}\xi_{\beta},\label{Gab}
\end{equation}
where $G_{\alpha\beta}$ with Greek subscripts $\alpha$, $\beta=\pm$
are components of the CTPGF matrix in single time representation in
Eq.(\ref{hatG}).

\subsection{Generating functionals of CTPGFs}

We begin with the Lagrangian density of the real boson field as
follows
\begin{eqnarray}
\mathcal{L}&=&\frac{1}{2}\partial_{\mu}\varphi(x)\partial^{\mu}\varphi(x)-\frac{1}{2}m^{2}\varphi^{2}(x)+\mathcal{L}_{int}(\varphi(x))\nonumber\\
&=&\mathcal{L}_{0}(\varphi(x))+\mathcal{L}_{int}(\varphi(x)),\label{}
\end{eqnarray}
where $\mathcal{L}_{0}$ is the free field term and
$\mathcal{L}_{int}$ is the interaction term. The generating
functional for the CTPGFs is defined as
\begin{equation}
Z[J(x)]\equiv\mathrm{Tr}\Big\{T_{p}\Big[\exp\Big(i\int_{p}d^{4}x
J(x)\varphi(x)\Big)\Big]\hat{\rho}\Big\},\label{ZJ1}
\end{equation}
where the time integration is
\begin{eqnarray}
\int_{p}dt&=&\int_{-\infty}^{+\infty}dt_{+}+\int_{+\infty}^{-\infty}dt_{-}\nonumber\\
&=&\int_{-\infty}^{+\infty}dt_{+}-\int_{-\infty}^{+\infty}dt_{-}.\label{}
\end{eqnarray}
The $n$-point CTPGF is defined as
\begin{eqnarray}
G_{p}(1\cdots n)&\equiv&
(-i)^{n-1}\mathrm{Tr}\Big[T_{p}(\varphi(1)\cdots\varphi(n))\hat{\rho}\Big]\nonumber\\
&=&i(-1)^{n}\frac{\delta^{n}Z[J(x)]}{\delta J(1)\cdots \delta
J(n)}\Big|_{J=0}.\label{}
\end{eqnarray}
In the interaction picture the generating functional $Z[J(x)]$ in
Eq.(\ref{ZJ1}) can be expressed as
\begin{eqnarray}
Z[J(x)]&=&\mathrm{Tr}\Big\{T_{p}\Big[\exp\Big(i\int_{p}d^{4}x
\big(\mathcal{L}_{int}(\varphi_{I}(x))+J(x)\varphi_{I}(x)\big)\Big)\Big]\hat{\rho}\Big\}\nonumber\\
&=&\exp\Big[i\int_{p}d^{4}x\mathcal{L}_{int}\Big(-i\frac{\delta}{\delta
J(x)}\Big)\Big]\nonumber\\
&\times&\mathrm{Tr}\Big\{T_{p}\Big[\exp\Big(i\int_{p}d^{4}x
J(x)\varphi_{I}(x)\Big)\Big]\hat{\rho}\Big\}.\label{ZJ2}
\end{eqnarray}
Using the Generalized Wick theorem~\cite{Hall1975}, one can
show~\cite{Zhou1981,Zhou1982,Chou1985}
\begin{equation}
T_{p}\Big[\exp\Big(i\int_{p}d^{4}x
J(x)\varphi_{I}(x)\Big)\Big]=Z_{0}[J(x)]:\exp\Big[i\int_{p}d^{4}x
J(x)\varphi_{I}(x)\Big]\!\!:\, ,\label{Tp}
\end{equation}
with
\begin{equation}
Z_{0}[J(x)]=\int_{p}[d
\varphi(x)]\exp\Big[i\int_{p}d^{4}x\Big(\mathcal{L}_{0}(\varphi(x))+J(x)\varphi(x)\Big)\Big].\label{}
\end{equation}
Here $:\,:$ denotes the normal product. Substituting Eq.(\ref{Tp})
into Eq.(\ref{ZJ2}), one obtains
\begin{equation}
Z[J(x)]=\exp\Big[i\int_{p}d^{4}x\mathcal{L}_{int}\Big(-i\frac{\delta}{\delta
J(x)}\Big)\Big]Z_{0}[J(x)] N[J(x)],\label{ZJ3}
\end{equation}
here
\begin{eqnarray}
N[J(x)]&=&\mathrm{Tr}\Big\{:\exp\Big[i\int_{p}d^{4}x
J(x)\varphi_{I}(x)\Big]\!\!:\hat{\rho}\Big\}\nonumber\\
&\equiv&\exp(iW_{p}^{N}[J(x)])\label{}
\end{eqnarray}
with
\begin{equation}
W_{p}^{N}[J(x)]=\sum_{n=1}^{\infty}\frac{i^{n-1}}{n!}\int_{p}d1\cdots\int_{p}dn
J(1)\cdots
J(n)\mathrm{Tr}[:\varphi_{I}(1)\cdots\varphi_{I}(n)\!\!:\hat{\rho}]_{c},\label{}
\end{equation}
where $N[J(x)]$ is the correlation functional for the initial state.

After some calculations, the generating functional $Z[J(x)]$ in
Eq.(\ref{ZJ3}) can be expressed as~\cite{Zhou1982}
\begin{eqnarray}
Z[J(x)]&=&\int_{p}[d
\varphi(x)]\exp\Big[i\int_{p}d^{4}x\Big(\mathcal{L}_{0}(\varphi(x))+\mathcal{L}_{int}(\varphi(x))+J(x)\varphi(x)\Big)\Big]
\nonumber\\&\times& \exp\Big[iW_{p}^{N}\Big(-\int_{p}d^{4}y
G^{-1}_{0}(x,y)\varphi(y)\Big)\Big],\label{ZJ4}
\end{eqnarray}
where $G^{-1}_{0}$ is the inverse of the propagator of the real
boson field, given by
\begin{equation}
G^{-1}_{0}(x,y)=(-\partial_{\mu}\partial^{\mu}-m^{2})\delta_{p}(x-y).\label{}
\end{equation}
In the single time presentation $\delta_{p}(x-y)$ can be expressed
as
\begin{equation}
\delta_{p}(x-y)=\delta(x-y)\sigma_{3}.\label{}
\end{equation}

The generating functional for the connected CTPGFs is given by
\begin{equation}
W[J(x)]=-i\ln Z[J(x)].\label{}
\end{equation}
Correspondingly, one can obtain
\begin{eqnarray}
G_{p}^{c}(1\cdots n)&\equiv&
(-i)^{n-1}\mathrm{Tr}\Big[T_{p}(\varphi(1)\cdots\varphi(n))\hat{\rho}\Big]_{c}\nonumber\\
&=&(-1)^{n-1}\frac{\delta^{n}W[J(x)]}{\delta J(1)\cdots \delta
J(n)}\Big|_{J=0}.\label{}
\end{eqnarray}
For $n=1$ we have
\begin{equation}
\langle\varphi(x)\rangle=\frac{\delta W[J(x)]}{\delta J(x)}.\label{}
\end{equation}
The vertex generating functional can be obtained by performing the
Legendre transformation as follows
\begin{equation}
\Gamma[\langle\varphi(x)\rangle]=W[J(x)]-\int_{p}d^{4}x
J(x)\langle\varphi(x)\rangle.\label{}
\end{equation}
It can be easily found that
\begin{equation}
\frac{\delta \Gamma[\langle\varphi(x)\rangle]}{\delta
\langle\varphi(x)\rangle}=-J(x).\label{}
\end{equation}

Next, we express the external source term in the generating
functional in Eq.(\ref{ZJ4}) in the single time representation as
\begin{eqnarray}
I_{s}&=&
\int_{p}d^{4}x J(x)\varphi(x)\nonumber\\
&=&\int_{-\infty}^{+\infty}dtd^{3}x \big(J_{+}(x)\varphi_{+}(x)-J_{-}(x)\varphi_{-}(x)\big)\nonumber\\
&=&\int d^{4}x \hat{J}^{\dag}(x)\sigma_{3}\hat{\varphi}(x),\label{}
\end{eqnarray}
with
\begin{equation}
\hat{\varphi}(x)=\Bigg(\begin{array}{c} \varphi_{+}(x)\\
\varphi_{-}(x) \end{array}\Bigg),\qquad \hat{J}(x)=\Bigg(\begin{array}{c} J_{+}(x) \\
J_{-}(x)\end{array}\Bigg).\label{}
\end{equation}
Introducing
\begin{eqnarray}
J_{\vartriangle}(x)&\equiv&J_{+}(x)-J_{-}(x),\qquad
J_{c}(x)\equiv\frac{1}{2}(J_{+}(x)+J_{-}(x)),
\nonumber\\
\varphi_{\vartriangle}(x)&\equiv&
\varphi_{+}(x)-\varphi_{-}(x),\qquad
\varphi_{c}(x)\equiv\frac{1}{2}(\varphi_{+}(x)+\varphi_{-}(x)),
\label{}
\end{eqnarray}
one can express $I_{s}$ as
\begin{equation}
I_{s}=\int d^{4}x
(J_{\vartriangle}(x)\varphi_{c}(x)+J_{c}(x)\varphi_{\vartriangle}(x)).\label{}
\end{equation}
Then we can obtain the generating functional for the CTPGFs in the
physical presentation as follows
\begin{equation}
Z[J_{\vartriangle}(x),J_{c}(x)]=Z[J_{+}(x),J_{-}(x)].\label{}
\end{equation}
In the same way, we have
\begin{equation}
W[J_{\vartriangle}(x),J_{c}(x)]=-i\ln
Z[J_{\vartriangle}(x),J_{c}(x)].\label{}
\end{equation}
and
\begin{eqnarray}
\Gamma[\langle\varphi_{\vartriangle}(x)\rangle,\langle\varphi_{c}(x)\rangle]
&=&W[J_{\vartriangle}(x),J_{c}(x)]\nonumber\\
&-&\int d^{4}x
\big(J_{\vartriangle}(x)\langle\varphi_{c}(x)\rangle+J_{c}(x)\langle\varphi_{\vartriangle}(x)\rangle\big),\label{}
\end{eqnarray}
with
\begin{equation}
\langle\varphi_{c}(x)\rangle=\frac{\delta
W[J_{\vartriangle}(x),J_{c}(x)]}{\delta J_{\vartriangle}(x)},\qquad
\langle\varphi_{\vartriangle}(x)\rangle=\frac{\delta
W[J_{\vartriangle}(x),J_{c}(x)]}{\delta J_{c}(x)}.\label{}
\end{equation}
Therefore, one can also obtain
\begin{equation}
\frac{\delta
\Gamma[\langle\varphi_{\vartriangle}(x)\rangle,\langle\varphi_{c}(x)\rangle]}{\delta
\langle\varphi_{\vartriangle}(x)\rangle}=-J_{c}(x),\qquad
\frac{\delta
\Gamma[\langle\varphi_{\vartriangle}(x)\rangle,\langle\varphi_{c}(x)\rangle]}{\delta
\langle\varphi_{c}(x)\rangle}=-J_{\vartriangle}(x).\label{Eq44}
\end{equation}

\section{Application of the CTPGFs in the NJL model}
\label{Sec3}

In this section we use the CTPGFs to study the spacially and
temporally inhomogeneous quark matter in the NJL model. The
Lagrangian density for the two-flavor NJL model is given
by~\cite{Nambu1961,Volkov1984,Klevansky1992,Hatsuda1994,Alkofer1996,Buballa2005}
\begin{eqnarray}
\mathcal{L}_{NJL}&=&\bar{\psi}(i\gamma_{\mu}\partial^{\mu}-\hat{m}_{0})\psi
 +G\Big[\big(\bar{\psi}\psi\big)^{2}
 +\big(\bar{\psi}i\gamma_{5}\vec{\tau}\psi\big)^{2}\Big],\label{lagrangian}
\end{eqnarray}
where $\psi=(\psi_{u},\psi_{d})^{T}$ is the quark field, and
$\hat{m}_{0}=\textrm{diag}(m_{u},m_{d})$ is the current quark mass
matrix. Throughout this work, we take $m_{u}=m_{d}\equiv m_{0}$,
assuming that the isospin symmetry is reserved on the Lagrangian
level. The four-fermion interaction with an effective coupling
strength $G$ for the scalar and pseudoscalar channels has
$\mathrm{SU_{V}}(2)\times \mathrm{SU_{A}}(2)\times
\mathrm{U_{V}}(1)$ symmetry, which is broken to
$\mathrm{SU_{V}}(2)\times \mathrm{U_{V}}(1)$ when $m_{0}\neq 0$.
Here $\tau^{a}(a=1,2,3)$ in the Lagrangian density in
Eq.(\ref{lagrangian}) are Pauli matrices in flavor space.

Following the CTPGF formalism in Sec.~\ref{Sec2}, we can obtain the
generating functional for the NJL model as follows
\begin{eqnarray}
Z[h,h^{a};\bar{\eta},\eta]
&=&\int_{p}[d\bar{\psi}][d\psi]\exp\Big\{i\int_{p}d^{4}x\Big(\mathcal{L}_{0}(\bar{\psi},\psi)+\mathcal{L}_{int}(\bar{\psi},\psi)+\bar{\eta}\psi+\bar{\psi}\eta\nonumber\\
&+&h\bar{\psi}\psi+h^{a}\bar{\psi}i\gamma_{5}\tau^{a}\psi+W_{p}^{N}[-\bar{\psi}
\overleftarrow{S}_{0}^{-1},
-\overrightarrow{S}_{0}^{-1}\psi]\Big)\Big\},\label{Znjl}
\end{eqnarray}
with
\begin{eqnarray}
\mathcal{L}_{0}(\bar{\psi},\psi)
&=&\bar{\psi}(i\gamma_{\mu}\partial^{\mu}-\hat{m}_{0})\psi,\nonumber\\
\mathcal{L}_{int}(\bar{\psi},\psi)&=&G\Big[\big(\bar{\psi}\psi\big)^{2}
 +\big(\bar{\psi}i\gamma_{5}\vec{\tau}\psi\big)^{2}\Big],\label{}
\end{eqnarray}
where $\bar{\eta}$, $\eta$ are the external sources for field $\psi$
and $\bar{\psi}$, respectively; $h$, $h^{a}$ for composite operators
$\bar{\psi}\psi$ and $\bar{\psi}i\gamma_{5}\tau^{a}\psi$. The term
$W_{p}^{N}$ in Eq.(\ref{Znjl}) is due to initial correlations as
shown in the last section, and here $S_{0}^{-1}$ is the inverse of
the propagator for the free fermion field. Up to a constant, we have
\begin{eqnarray}
&&\exp\Big\{i\int_{p}d^{4}x\Big(G\Big[\big(\bar{\psi}\psi\big)^{2}
 +\big(\bar{\psi}i\gamma_{5}\vec{\tau}\psi\big)^{2}\Big]+h\bar{\psi}\psi+h^{a}\bar{\psi}i\gamma_{5}\tau^{a}\psi\Big)\Big\}\nonumber\\
&=&\int_{p}[d\sigma][d\pi^{a}]\exp\Big\{i\int_{p}d^{4}x\Big(-G(\sigma^{2}+{\pi^{a}}^{2})+(2G\bar{\psi}\psi+h)\sigma
\nonumber\\
&+&(2G\bar{\psi}i\gamma_{5}\tau^{a}\psi+h^{a})\pi^{a}-\frac{1}{4G}(h^{2}+{h^{a}}^{2})\Big)\Big\}
.\label{Eq48}
\end{eqnarray}
Substituting Eq.(\ref{Eq48}) into Eq.(\ref{Znjl}), we obtain
\begin{eqnarray}
Z[h,h^{a};\bar{\eta},\eta]
&=&\exp\big[-\frac{i}{4G}\int_{p}d^{4}x(h^{2}+{h^{a}}^{2})\big]\int_{p}[d\sigma][d\pi^{a}]\int_{p}[d\bar{\psi}][d\psi]\nonumber\\
&\times&\exp\Big\{i\int_{p}d^{4}x\Big(\bar{\psi}S^{-1}\psi-G(\sigma^{2}+{\pi^{a}}^{2})
+h\sigma+h^{a}\pi^{a}\nonumber\\
&+&\bar{\eta}\psi+\bar{\psi}\eta+W_{p}^{N}[-\bar{\psi}
\overleftarrow{S}_{0}^{-1},
-\overrightarrow{S}_{0}^{-1}\psi]\Big)\Big\},\label{Znj2}
\end{eqnarray}
with
\begin{equation}
S^{-1}(x,y)=\big[i\gamma_{\mu}\partial^{\mu}-(\hat{m}_{0}-2G\sigma)+2Gi\gamma_{5}\pi^{a}\tau^{a}\big]\delta_{p}(x-y).\label{Eq50}
\end{equation}
In the realistic calculations for the equilibrium (or local
equilibrium) system, the contribution of the density matrix, i.e.,
the term $W_{p}^{N}$ in Eq.(\ref{Znj2}) can be neglected temporarily
for simplicity, and its contribution can be recovered through the
fluctuation-dissipation theorem satisfied by the CTPGFs. More
detailed discussions can be found in
Refs.~\cite{Chou1985,Zhou1982c}. Neglecting $W_{p}^{N}$ in
Eq.(\ref{Znj2}) temporarily, we can integrate over the fermion field
and obtain
\begin{eqnarray}
Z[h,h^{a};\bar{\eta},\eta]
&=&\exp\big[-\frac{i}{4G}\int_{p}d^{4}x(h^{2}+{h^{a}}^{2})\big]\int_{p}[d\sigma][d\pi^{a}]\nonumber\\
&\times&\exp\Big\{i\Big[
I_{0}(\sigma,\pi^{a})+I_{\mathrm{det}}(\sigma,\pi^{a})\nonumber\\
&+&\int_{p}d^{4}x(h\sigma+h^{a}\pi^{a})-\int_{p}d^{4}x
d^{4}y\bar{\eta}(x)S(x,y)\eta(y)\Big]\Big\},\label{}
\end{eqnarray}
with
\begin{eqnarray}
I_{0}(\sigma,\pi^{a})
&=&-\int_{p}d^{4}x G(\sigma^{2}+{\pi^{a}}^{2}),\nonumber\\
I_{\mathrm{det}}(\sigma,\pi^{a})&=&-i\mathrm{Tr}\ln S^{-1},\label{}
\end{eqnarray}
where $\mathrm{Tr}$ means trace operation in both the inner space
(Dirac, flavor, and color) and the coordinate space. Then we can
introduce the effective action for the $\sigma$ and $\pi$ fields as
follows
\begin{equation}
I^{\mathrm{eff}}(\sigma,\pi^{a})=I_{0}(\sigma,\pi^{a})+I_{\mathrm{det}}(\sigma,\pi^{a}).\label{}
\end{equation}
In the mean field approximation, the vertex generating functional
can be approximated by the effective action~\cite{Jackiw1974}, i.e.,
\begin{equation}
\Gamma(\sigma,\pi^{a})\simeq
I^{\mathrm{eff}}(\sigma,\pi^{a}),\label{Eq54}
\end{equation}
where $\sigma$ and $\pi^{a}$ mean $\langle\sigma\rangle$ and
$\langle\pi^{a}\rangle$, without confusions. In the physical
representation we have
\begin{eqnarray}
\Gamma(\sigma_{\vartriangle},\pi_{\vartriangle}^{a};
\sigma_{c},\pi_{c}^{a})
&\simeq&I^{\mathrm{eff}}(\sigma_{\vartriangle},\pi_{\vartriangle}^{a};
\sigma_{c},\pi_{c}^{a})\nonumber\\
&=&I_{0}(\sigma_{\vartriangle},\pi_{\vartriangle}^{a};
\sigma_{c},\pi_{c}^{a})+I_{\mathrm{det}}(\sigma_{\vartriangle},\pi_{\vartriangle}^{a};
\sigma_{c},\pi_{c}^{a}),\label{}
\end{eqnarray}
with
\begin{equation}
I_{0}(\sigma_{\vartriangle},\pi_{\vartriangle}^{a};
\sigma_{c},\pi_{c}^{a})=-2G\int
d^{4}x(\sigma_{\vartriangle}\sigma_{c}+\pi_{\vartriangle}^{a}\pi_{c}^{a}),\label{}
\end{equation}
and
\begin{eqnarray}
[S^{-1}(x,y)]_{\alpha\beta}&=&\Big[(i\gamma_{\mu}\partial^{\mu}-\hat{m}_{0})\xi_{\alpha}\eta_{\beta}+2G(\xi_{\alpha}\sigma_{c}
+\frac{1}{2}\eta_{\alpha}\sigma_{\vartriangle})\eta_{\beta}\nonumber\\
&+&2Gi\gamma_{5}\tau^{a}(\xi_{\alpha}\pi_{c}^{a}
+\frac{1}{2}\eta_{\alpha}\pi_{\vartriangle}^{a})\eta_{\beta}\Big]\delta_{\alpha\beta}\delta^{4}(x-y).\label{Eq57}
\end{eqnarray}
Following the notations in the last section, we have
\begin{eqnarray}
\sigma_{\vartriangle}(x)&=&\sigma_{+}(x)-\sigma_{-}(x),\qquad
\sigma_{c}(x)=\frac{1}{2}(\sigma_{+}(x)+\sigma_{-}(x)),\nonumber\\
\pi_{\vartriangle}^{a}(x)&=&\pi_{+}^{a}(x)-\pi_{-}^{a}(x),\qquad
\pi_{c}^{a}(x)=\frac{1}{2}(\pi_{+}^{a}(x)+\pi_{-}^{a}(x)).\label{}
\end{eqnarray}
In the following, we neglect the influence of the $\pi^{a}$ field
and concentrate on the order parameter of chiral phase transition,
i.e., the $\sigma$ field. First of all, let us consider the simplest
stationary homogeneous quark system, i.e., the $\sigma$ field is
independent of the time and space coordinates. From Eq.(\ref{Eq44})
and Eq.(\ref{Eq54}) we have
\begin{equation}
\frac{\delta I^{\mathrm{eff}}(\sigma_{\vartriangle},
\sigma_{c})}{\delta
\sigma_{\vartriangle}}\Big|_{\sigma_{\vartriangle}=0,
\sigma_{c}=\sigma}=-h_{c}=0,\label{Eq59}
\end{equation}
For the left hand side of Eq.(\ref{Eq59}), we have
\begin{equation}
\frac{\delta I_{0}(\sigma_{\vartriangle}, \sigma_{c})}{\delta
\sigma_{\vartriangle}}=-2G\sigma_{c},\label{Eq60}
\end{equation}
and
\begin{equation}
\frac{\delta I_{\mathrm{det}}(\sigma_{\vartriangle},
\sigma_{c})}{\delta
\sigma_{\vartriangle}}=-i\mathrm{Tr}\Big(S\frac{\delta
S^{-1}}{\delta \sigma_{\vartriangle}}\Big).\label{}
\end{equation}
From Eq.(\ref{Eq57}), we have
\begin{equation}
\frac{\delta [S^{-1}(y,z)]_{\alpha\beta}}{\delta
\sigma_{\vartriangle}(x)}=G\delta_{\alpha\beta}\delta^{4}(y-x)\delta^{4}(y-z).\label{Eq62}
\end{equation}
Then
\begin{eqnarray}
\frac{\delta I_{\mathrm{det}}(\sigma_{\vartriangle},
\sigma_{c})}{\delta \sigma_{\vartriangle}(x)}&=&-i\int_{p}d^{4}y
d^{4}z\,\mathrm{tr}\Big[S(y,z)\frac{\delta S^{-1}(y,z)}{\delta
\sigma_{\vartriangle}(x)}\Big],\nonumber\\
&=&-i\int d^{4}y
d^{4}z\,\mathrm{tr}\Big[\sigma_{3}^{p}S(y,z)\sigma_{3}^{p}\frac{\delta
S^{-1}(y,z)}{\delta \sigma_{\vartriangle}(x)}\Big]\nonumber\\
&=&-i\int d^{4}y d^{4}z\,\mathrm{tr}\Big[S(y,z)\frac{\delta
S^{-1}(y,z)}{\delta \sigma_{\vartriangle}(x)}\Big],\label{}
\end{eqnarray}
where $\mathrm{tr}$ means trace operation not including in the
coordinate space; $\sigma_{3}^{p}$ is the Pauli matrix in the single
time representation. Substituting Eq.(\ref{Gab}) and Eq.(\ref{Eq62})
into above equation, we obtain
\begin{equation}
\frac{\delta I_{\mathrm{det}}(\sigma_{\vartriangle},
\sigma_{c})}{\delta
\sigma_{\vartriangle}(x)}=-iG\mathrm{tr}S_{c}(x,x).\label{Eq64}
\end{equation}
For the inverse of the quark propagator in Eq.(\ref{Eq50}) without
$\pi^{a}$ field, we can easily obtain
\begin{eqnarray}
S_{r}(p)-S_{a}(p)&=&S_{-}(p)-S_{+}(p),\nonumber\\
&=&-i\pi\frac{\slash\!\!\!p+M}{E_{p}}[\delta(p_{0}-E_{p})-\delta(p_{0}+E_{p})],\label{}
\end{eqnarray}
in the momentum space. Here we have
\begin{equation}
M=m_{0}-2G\sigma, \qquad E_{p}=\sqrt{p^{2}+M^{2}}.\label{Eq66}
\end{equation}
Employing the fluctuation-dissipation theorem~\cite{Chou1985}, we
obtain
\begin{eqnarray}
S_{c}(p)&=&\tanh\big[\frac{\beta}{2}(p_{0}-\mu)\big]\big[S_{r}(p)-S_{a}(p)\big],\nonumber\\
&=&-i\pi\frac{\slash\!\!\!p+M}{E_{p}}[\delta(p_{0}-E_{p})-\delta(p_{0}+E_{p})]\tanh\big[\frac{\beta}{2}(p_{0}-\mu)\big],\label{Eq67}
\end{eqnarray}
where $\beta=1/T$ is the inverse of the temperature, $\mu$ the quark
chemical potential. Substituting Eq.(\ref{Eq67}) into
Eq.(\ref{Eq64}), we obtain
\begin{eqnarray}
\frac{\delta I_{\mathrm{det}}(\sigma_{\vartriangle},
\sigma_{c})}{\delta
\sigma_{\vartriangle}(x)}&=&-iG\int\frac{d^{4}p}{(2\pi)^{4}}\mathrm{tr}S_{c}(p),\nonumber\\
&=&-4GN_{f}N_{c}\int\frac{d^{3}p}{(2\pi)^{3}}\frac{M}{E_{p}}\Big[1-\frac{1}{e^{\beta(E_{p}-\mu)}+1}\nonumber\\
&-&\frac{1}{e^{\beta(E_{p}+\mu)}+1}\Big],\label{Eq68}
\end{eqnarray}
where $N_{f}$ is the number of quark flavor, $N_{c}$ the color
number. Combining Eqs.(\ref{Eq59})(\ref{Eq60}) with Eq.(\ref{Eq68}),
we obtain
\begin{equation}
\sigma=-2N_{f}N_{c}\int\frac{d^{3}p}{(2\pi)^{3}}\frac{M}{E_{p}}\Big[1-\frac{1}{e^{\beta(E_{p}-\mu)}+1}-\frac{1}{e^{\beta(E_{p}+\mu)}+1}\Big],\label{}
\end{equation}
which is the gap equation for the order parameter in the NJL model.

In the following we will consider the weak non-stationary
inhomogeneous quark matter, where the ``weak'' means that we can
expand the left hand side of Eq.(\ref{Eq59}) as power series of the
gradient of the $\sigma(t,\,\vec{x})$ field. The efficiency of this
expansion has be verified in Ref.~\cite{Zhou1982c}. Furthermore, we
only concentrate on the quark system which is near the chiral phase
transition, therefore, the order parameter, i.e., the expectation
value of the $\sigma(t,\,\vec{x})$ field is smaller than the
critical temperature. Then, the left hand side of Eq.(\ref{Eq59})
can be expressed as
\begin{eqnarray}
\frac{\delta I^{\mathrm{eff}}(\sigma_{\vartriangle},
\sigma_{c})}{\delta
\sigma_{\vartriangle}(x)}\Big|_{\sigma_{\vartriangle}=0,
\sigma_{c}=\sigma(x)}&\simeq&\frac{\delta
I^{\mathrm{eff}}(\sigma_{\vartriangle}, \sigma_{c})}{\delta
\sigma_{\vartriangle}(x)}\Big|_{\sigma_{\vartriangle}=0,
\sigma_{c}=\sigma}^{'}\nonumber\\
&+&\int
d^{4}y\frac{\delta^{2}I^{\mathrm{eff}}(\sigma_{\vartriangle},
\sigma_{c})}{\delta\sigma_{\vartriangle}(x+\frac{y}{2})\delta\sigma_{c}(x-\frac{y}{2})}\Big[\sigma(x-\frac{y}{2})\nonumber\\
&-&\sigma(x+\frac{y}{2})\Big],\label{Eq70}
\end{eqnarray}
where the dependence of the $\sigma$ field in the fermion propagator
on the coordinate is neglected in the first term of the right hand
side of Eq.(\ref{Eq70}), i.e., it is just the result for the
homogeneous system with $\sigma$ replaced by $\sigma(x)$. The second
term of the right hand side of Eq.(\ref{Eq70}) can be expressed as
\begin{eqnarray}
&&\int d^{4}y\frac{\delta^{2}I^{\mathrm{eff}}(\sigma_{\vartriangle},
\sigma_{c})}{\delta\sigma_{\vartriangle}(x+\frac{y}{2})\delta\sigma_{c}(x-\frac{y}{2})}\Big[\sigma(x-\frac{y}{2})-\sigma(x+\frac{y}{2})\Big]
\nonumber\\
&\simeq&\int
d^{4}y\frac{\delta^{2}I^{\mathrm{eff}}(\sigma_{\vartriangle},
\sigma_{c})}{\delta\sigma_{\vartriangle}(x+\frac{y}{2})\delta\sigma_{c}(x-\frac{y}{2})}\Big[-y^{\mu}\partial_{\mu}\sigma(x)+\frac{1}{2}y^{\mu}y^{\nu}
\partial_{\mu}\partial_{\nu}\sigma(x)\Big]\nonumber\\
&=&\int
d^{4}y\frac{\delta^{2}I^{\mathrm{eff}}(\sigma_{\vartriangle},
\sigma_{c})}{\delta\sigma_{\vartriangle}(x+\frac{y}{2})\delta\sigma_{c}(x-\frac{y}{2})}\Big[
-\Big(-i\frac{\partial}{\partial
q_{\mu}}e^{iqy}\Big)_{q=0}\partial_{\mu}\sigma(x)\nonumber\\
&+&\frac{1}{2}(-i)^{2}\Big(\frac{\partial^{2}}{\partial
q_{\mu}\partial
q_{\nu}}e^{iqy}\Big)_{q=0}\partial_{\mu}\partial_{\nu}\sigma(x)\Big]\nonumber\\
&=&i\Big[\frac{\partial}{\partial
q_{\mu}}\Gamma_{r}(x,q)\Big]_{q=0}\partial_{\mu}\sigma(x)-\frac{1}{2}\Big[\frac{\partial^{2}}{\partial
q_{\mu}\partial
q_{\nu}}\Gamma_{r}(x,q)\Big]_{q=0}\partial_{\mu}\partial_{\nu}\sigma(x),\label{Eq71}
\end{eqnarray}
with
\begin{equation}
\Gamma_{r}(x,q)=\int
d^{4}y\frac{\delta^{2}I^{\mathrm{eff}}(\sigma_{\vartriangle},
\sigma_{c})}{\delta\sigma_{\vartriangle}(x+\frac{y}{2})\delta\sigma_{c}(x-\frac{y}{2})}e^{iqy},\label{}
\end{equation}
where $\mu,\nu=$0, 1, 2, 3. We have
\begin{equation}
\frac{\delta^{2}I^{\mathrm{eff}}(\sigma_{\vartriangle},
\sigma_{c})}{\delta\sigma_{\vartriangle}(x+\frac{y}{2})\delta\sigma_{c}(x-\frac{y}{2})}
=\frac{\delta^{2}I_{0}(\sigma_{\vartriangle},
\sigma_{c})}{\delta\sigma_{\vartriangle}(x+\frac{y}{2})\delta\sigma_{c}(x-\frac{y}{2})}
+\frac{\delta^{2}I_{\mathrm{det}}(\sigma_{\vartriangle},
\sigma_{c})}{\delta\sigma_{\vartriangle}(x+\frac{y}{2})\delta\sigma_{c}(x-\frac{y}{2})},\label{}
\end{equation}
with
\begin{equation}
\frac{\delta^{2}I_{0}(\sigma_{\vartriangle},
\sigma_{c})}{\delta\sigma_{\vartriangle}(x+\frac{y}{2})\delta\sigma_{c}(x-\frac{y}{2})}=-2G\delta^{4}(y),\label{}
\end{equation}
and
\begin{equation}
\frac{\delta^{2}I_{\mathrm{det}}(\sigma_{\vartriangle},
\sigma_{c})}{\delta\sigma_{\vartriangle}(x+\frac{y}{2})\delta\sigma_{c}(x-\frac{y}{2})}=i\mathrm{Tr}\Big[
S\frac{\delta
S^{-1}}{\delta\sigma_{\vartriangle}(x+\frac{y}{2})}S\frac{\delta
S^{-1}}{\delta\sigma_{c}(x-\frac{y}{2})}\Big].\label{}
\end{equation}
After some calculations, we find
\begin{equation}
\Gamma_{r}(x,q)=-2G+i2G^{2}\int\frac{d^{4}p}{(2\pi)^{4}}\mathrm{tr}\big[
S_{c}(x,p)S_{r}(x,p+q)+S_{a}(x,p)S_{c}(x,p+q)\big].\label{Eq76}
\end{equation}
The correlation Green function for quark field $S_{c}$ is given in
Eq.(\ref{Eq67}), and the retarded and the advanced Green functions
are given by
\begin{eqnarray}
S_{r}(p)&=&\frac{1}{\slash\!\!\!p-M+i\epsilon}+i\pi\frac{\slash\!\!\!p+M}{E_{p}}\delta(p_{0}+E_{p}),\nonumber\\
S_{a}(p)&=&\frac{1}{\slash\!\!\!p-M+i\epsilon}+i\pi\frac{\slash\!\!\!p+M}{E_{p}}\delta(p_{0}-E_{p}).\label{}
\end{eqnarray}
Here, the dependence of the Green functions on the space and time
coordinates is realized through the $\sigma(x)$ field as shown in
Eq.(\ref{Eq66}). Performing calculations on Eq.(\ref{Eq76}), we
obtain
\begin{eqnarray}
\Gamma_{r}(x,q)&=&-2G\nonumber\\
&+&4G^{2}N_{c}N_{f}\int\frac{d^{3}p}{(2\pi)^{3}}\Big(
\tanh\big[\frac{\beta}{2}(E_{p}-\mu)\big]+\tanh\big[\frac{\beta}{2}(E_{p}+\mu)\big]\Big)\nonumber\\
&\times&\frac{1}{E_p}\Big(\frac{2M^2+E_{p}q_0-\vec{p}\cdot\vec{q}}{(q_0+E_p)^2-E^2_{p+q}+i\epsilon(q_0+E_p)}\nonumber\\
&+&
\frac{2M^2-E_{p}q_0-\vec{p}\cdot\vec{q}}{(q_0-E_p)^2-E^2_{p+q}+i\epsilon(q_0-E_p)}\Big),\label{GenEq}
\end{eqnarray}
When considering the variation with respect to the space
coordinates, we can simplify Eq.(\ref{GenEq}) further by taking the
limit $q_0\rightarrow 0$ firstly, which gives
\begin{eqnarray}
\Gamma_{r}(x,q)_{q_{0}=0}&=&-2G\nonumber\\
&+&4G^{2}N_{c}N_{f}\int\frac{d^{3}p}{(2\pi)^{3}}\Big(
\tanh\big[\frac{\beta}{2}(E_{p}-\mu)\big]+\tanh\big[\frac{\beta}{2}(E_{p}+\mu)\big]\Big)\nonumber\\
&\times&\frac{2M^{2}-\vec{p}\cdot\vec{q}}{E_{p}E_{p+q}}\Big(\mathcal{P}\frac{1}{E_{p}-E_{p+q}}
-\mathcal{P}\frac{1}{E_{p}+E_{p+q}}\Big),\label{Eq78}
\end{eqnarray}
where $\mathcal{P}$ denotes main value, and
\begin{equation}
E_{p+q}=[(\vec{p}+\vec{q})^{2}+M^{2}]^{1/2}.\label{}
\end{equation}
when $|\vec{q}|\rightarrow 0$, we can expand the left hand side of
Eq.(\ref{Eq78}) as power series of $|\vec{q}|$. Furthermore, taking
into consideration that the system is near the chiral phase
transition, so the quark mass $M$ is smaller than the critical
temperature, we obtain
\begin{eqnarray}
\Gamma_{r}(x,q)_{q_{0}=0}&\simeq&-2G\nonumber\\
&+&4G^{2}N_{c}N_{f}\frac{1}{\pi^{2}}\int_{0}^{\Lambda}
dp \frac{p^{2}}{E_{p}}\Big[1-\frac{1}{e^{\beta(E_{p}-\mu)}+1}-\frac{1}{e^{\beta(E_{p}+\mu)}+1}\Big]\nonumber\\
&-&|\vec{q}|^{2}G^{2}N_{c}N_{f}\frac{1}{\pi^{2}}\int_{0}^{\Lambda}dp\frac{1}{E_{p}}
\Big[1-\frac{1}{e^{\beta(E_{p}-\mu)}+1}-\frac{1}{e^{\beta(E_{p}+\mu)}+1}\Big],\nonumber\\\label{Eq80}
\end{eqnarray}
where $\Lambda$ is the cut-off of the momentum integration which is
a parameter in the NJL model. From Eq.(\ref{Eq80}) we can easily
obtain
\begin{eqnarray}
\Big[\frac{\partial}{\partial
q^{i}}\Gamma_{r}(x,q)\Big]_{q=0}&=&0,\label{Eq81}\\
\Big[\frac{\partial^{2}}{\partial q^{i}\partial
q^{j}}\Gamma_{r}(x,q)\Big]_{q=0}&=&-\delta_{ij}2G^{2}N_{c}N_{f}\frac{1}{\pi^{2}}\int_{0}^{\Lambda}dp\frac{1}{E_{p}}
\Big[1-\frac{1}{e^{\beta(E_{p}-\mu)}+1}\nonumber\\
&&-\frac{1}{e^{\beta(E_{p}+\mu)}+1}\Big],\label{Eq82}
\end{eqnarray}
Substituting Eqs.(\ref{Eq81})(\ref{Eq82}) into
Eqs.(\ref{Eq70})(\ref{Eq71}) and combining Eq.(\ref{Eq59}), we
obtain
\begin{eqnarray}
&&\sigma(x)+N_{c}N_{f}\frac{1}{\pi^{2}}\int_{0}^{\Lambda}dp
p^{2}\frac{M}{E_{p}}
\Big[1-\frac{1}{e^{\beta(E_{p}-\mu)}+1}-\frac{1}{e^{\beta(E_{p}+\mu)}+1}\Big]\nonumber\\
&&-\frac{G}{2}N_{c}N_{f}\frac{1}{\pi^{2}}\int_{0}^{\Lambda}dp\frac{1}{E_{p}}
\Big[1-\frac{1}{e^{\beta(E_{p}-\mu)}+1}-\frac{1}{e^{\beta(E_{p}+\mu)}+1}\Big]\nabla^{2}\sigma(x)\nonumber\\
&&=0,\label{sGL}
\end{eqnarray}
which is the Ginzburg-Landau equation for the inhomogeneous quark
matter in the NJL model.

We can also consider the time variation of the quark matter near the
critical point by the same method that we applied to the space
coordinates, which gives us the time-dependent Ginzburg-Landau
equation. Without influencing the final result, we can take the
limit $\vec{q}\rightarrow 0$ and Eq.(\ref{GenEq}) is simplified to
the following form,
\begin{eqnarray}
\Gamma_{r}(x,q)_{\vec{q}=0}&=&-2G\nonumber\\
&+&4G^{2}N_{c}N_{f}\int\frac{d^{3}p}{(2\pi)^{3}}\Big(
\tanh\big[\frac{\beta}{2}(E_{p}-\mu)\big]+\tanh\big[\frac{\beta}{2}(E_{p}+\mu)\big]\Big)\nonumber\\
&\times&\frac{1}{E_p}\Big(\frac{2M^2+E_{p}q_0}{(q_0+E_p)^2-E^2_{p}+i\epsilon(q_0+E_p)}\nonumber\\
&+&
\frac{2M^2-E_{p}q_0}{(q_0-E_p)^2-E^2_{p}+i\epsilon(q_0-E_p)}\Big).\label{time}
\end{eqnarray}
In the same way, we ignore the mass term because the system is near
the QCD critical point. By expanding Eq.(\ref{time}) with respect to
$q_0$, we obtain
\begin{eqnarray}
\Gamma_{r}(x,q)_{\vec{q}=0}&=&-2G\nonumber\\
&+&4G^{2}N_{c}N_{f}\frac{1}{\pi^2}\int_{0}^{\Lambda}dp\frac{p^2}{E_{p}}
\Big[1-\frac{1}{e^{\beta(E_{p}-\mu)}+1}-\frac{1}{e^{\beta(E_{p}+\mu)}+1}\Big]\nonumber\\
&+& q_0^2
G^{2}N_{c}N_{f}\frac{1}{\pi^2}\int_{0}^{\Lambda}dp\frac{p^2}{E^3_{p}}
\Big[1-\frac{1}{e^{\beta(E_{p}-\mu)}+1}-\frac{1}{e^{\beta(E_{p}+\mu)}+1}\Big].\nonumber\\
\end{eqnarray}
Differentiating $\Gamma_r(x,q)$ with respect to $q_0$, we have
\begin{eqnarray}
\Big[\frac{\partial}{\partial
q^{0}}\Gamma_{r}(x,q)\Big]_{q=0}&=&0,\label{t1}\\
\Big[\frac{\partial^{2}}{\partial q^{0}\partial
q^{0}}\Gamma_{r}(x,q)\Big]_{q=0}&=&
2G^{2}N_{c}N_{f}\frac{1}{\pi^{2}}\int_{0}^{\Lambda}dp\frac{p^2}{E^3_{p}}
\Big[1-\frac{1}{e^{\beta(E_{p}-\mu)}+1}\nonumber\\
&&-\frac{1}{e^{\beta(E_{p}+\mu)}+1}\Big]\label{t2},
\end{eqnarray}
Substituting Eqs.(\ref{t1})(\ref{t2}) into Eqs.(\ref{Eq70})
(\ref{Eq71}) and combining Eq.(\ref{sGL}), we obtain the
time-dependent Ginzburg-Landau equation:
\begin{eqnarray}
&&\sigma(x)+N_{c}N_{f}\frac{1}{\pi^{2}}\int_{0}^{\Lambda}dp
p^{2}\frac{M}{E_{p}}
\Big[1-\frac{1}{e^{\beta(E_{p}-\mu)}+1}-\frac{1}{e^{\beta(E_{p}+\mu)}+1}\Big]\nonumber\\
&&-\frac{G}{2}N_{c}N_{f}\frac{1}{\pi^{2}}\int_{0}^{\Lambda}dp\frac{1}{E_{p}}
\Big[1-\frac{1}{e^{\beta(E_{p}-\mu)}+1}-\frac{1}{e^{\beta(E_{p}+\mu)}+1}\Big]\nabla^{2}\sigma(x)\nonumber\\
&&+\frac{G}{2}N_{c}N_{f}\frac{1}{\pi^{2}}\int_{0}^{\Lambda}dp\frac{p^2}{E^3_{p}}
\Big[1-\frac{1}{e^{\beta(E_{p}-\mu)}+1}-\frac{1}{e^{\beta(E_{p}+\mu)}+1}\Big]\partial^2_t\sigma(x)\nonumber\\
&&=0.\label{TDGL}
\end{eqnarray}
Note that in the above time-dependent Ginzburg-Landau equation, the
time derivative is second order, meaning that the $\sigma$ mode is a
propagating one, which is different from the diffusive
time-dependent Ginzburg-Landau equation for the precursory mode in
the superconductor phenomena~\cite{Kunihiro:1995rb,Hatsuda:1985eb}.

\section{Summary and Discussions}
\label{Sec4}

CTPGF formalism is a powerful tool, because it can deal with not
only equilibrium problems but also nonequilibrium ones. When much
attention are paid on the critical dynamics of the QCD phase
transitions, especially the chiral phase transition, it is quite
natural for people to choose the CTPGF formalism. In this work, the
CTPGF formalism is applied in the NJL model. First of all, we simply
review the CTPGF formalism, mainly on the generating functional of
the CTPGFs. Then we use this formalism to obtain the well-known gap
equation for the quark condensate in stationary homogeneous system.
We have also used this formalism to obtain the Ginzburg-Landau
equation for the chiral order parameter in stationary inhomogeneous
system and the time-dependent Ginzburg-Landau equation describing
the critical dynamics of the chiral phase transition. In our derived
Ginzburg-Landau equation, there is no other parameters except for
those in the original NJL model.

In the present paper, we only derive the time-dependent
Ginzburg-Landau equation. However, to fully appreciate the critical
dynamics of the chiral phase transition, we also need to analyze the
equation numerically and make some comparison with the experimental
data. Furthermore, the present derivation is only the mean field
result and we ignore the effects of the interaction of $\sigma(x)$
mode with $\pi(x)$ meson. Related works are under progress and we
will report them elsewhere.

\section*{Acknowledgments}
The authors are grateful for Prof. Yue-Liang Wu for interesting
discussions. One of the authors (W. J. F.) also acknowledges
financial support from China Postdoctoral Science Foundation No.
20090460534.

%\appendix


%% References


\bibliographystyle{model1a-num-names}

\begin{thebibliography}{00}


\bibitem{Shuryak2004}
E.~V.~Shuryak, Prog. Part. Nucl. Phys.  53 (2004) 273.

\bibitem{Gyulassy2005}
M.~Gyulassy, L.~McLerran, Nucl. Phys. A 750 (2005) 30.

\bibitem{Shuryak2005}
E.~V.~Shuryak, Nucl. Phys. A 750 (2005) 64.

\bibitem{Arsene2005}
I.~Arsene \textit{et al}, Nucl. Phys. A 757 (2005) 1.

\bibitem{Back2005}
B.~B.~Back \textit{et al}, Nucl. Phys. A 757 (2005) 28.

\bibitem{Adams2005}
J.~Adams \textit{et al}, Nucl. Phys. A 757 (2005) 102.

\bibitem{Adcox2005}
K.~Adcox \textit{et al}, Nucl. Phys. A 757 (2005) 184.

\bibitem{Blaizot2007}
J.-P.~Blaizot, J. Phys. G 34 (2007) S243.

\bibitem{Weber2005} F. Weber,
Prog. Part. Nucl. Phys. 54 (2005) 193.

\bibitem{Alford2007} M. Alford, D. Blaschke, A. Drago, T. Kl\"{a}hn,
G. Pagliara, J. Shaffner-Bielich, Nature 445 (2007) E 7.

\bibitem{Alford2008}
M.~Alford, A.~Schmitt, K.~Rajagopal, T. Sch\"{a}fer, Rev. Mod. Phys.
80 (2008) 1455.

\bibitem{Fu2008b}
W.~J.~Fu, H.~Q.~Wei, Y.~X.~Liu, Phys. Rev. Lett. 101 (2008) 181102.

\bibitem{Fodor2002}
Z.~Fodor, S.~D. Katz, Phys. Lett. B 534 (2002) 87; {\it ibid}, J.
High Energy Phys. 0203 (2002) 014; Z.~Fodor, S.~D. Katz, K.~K.
Szabo, Phys. Lett. B 568 (2003) 73.

\bibitem{Ejiri2004}
S.~Ejiri, C.~R.~Allton, S.~J.~Hands, O.~Kaczmarek, F.~Karsch,
E.~Laermann, C.~Schmidt, Prog. Theor. Phys. Suppl. 153 (2004) 118.

\bibitem{Gavai2005}
R.~V.~Gavai, S.~Gupta, Phys. Rev. D 71 (2005) 114014.

\bibitem{deForcrand2003}
P.~de Forcrand, O.~Philipsen,
%``The QCD phase diagram for three degenerate flavors and small baryon
%density,''
Nucl. Phys. B 642 (2002) 290; B 673 (2003) 170; P. de Focrand, S.
Kratochvila, Nucl. Phys. B, Proc. Suppl. 153 (2006) 62.

\bibitem{Asakawa1989}
M.~Asakawa, K.~Yazaki, Nucl. Phys. A 504 (1989) 668.

\bibitem{Barducci1989}
A.~Barducci, R.~Casalbuoni, S.~De Curtis, R.~Gatto, G.~Pettini,
Phys. Lett. B 231 (1989) 463; Phys. Rev.  D 41 (1990) 1610.

\bibitem{Barducci1994}
A.~Barducci, R.~Casalbuoni, G.~Pettini, R.~Gatto, Phys. Rev. D 49
(1994) 426.

\bibitem{Berges1999}
J.~Berges, K.~Rajagopal, Nucl. Phys. B 538 (1999) 215.

\bibitem{Halasz1998}
M. A. Halasz, A. D. Jackson, R. E. Shrock,M. A. Stephanov, J. J. M.
Verbaarschot, Phys. Rev. D 58 (1998) 096007.

\bibitem{Scavenius2001}
O. Scavenius, A. Mocsy, I. N. Mishustin, D. H. Rischke, Phys. Rev. C
64 (2001) 045202.

\bibitem{Hatta2003}
Y. Hatta, T. Ikeda, Phys. Rev. D 67 (2003) 014028.

\bibitem{Barducci2005}
A. Barducci, R. Casalbuoni, G. Pettini, L. Ravagli, Phys. Rev. D 72
(2005) 056002.

\bibitem{Fu2008}
W.~J.~Fu, Z.~Zhang, Y.~X.~Liu, Phys. Rev.  D 77 (2008) 014006.

\bibitem{Stephanov2006}
M.~A.~Stephanov, PoS LAT2006 (2006) 024.

\bibitem{Mohanty2009}
B. Mohanty, Nucl. Phys.  A 830 (2009) 899c; T. Schuster, PoS  CPOD
2009 (2009) 029; G. Stefanek, PoS  CPOD 2009 (2009) 049.

\bibitem{Anticic2009}
T. Anticic \textit{et al.} (NA49 Collaboration), N. G. Antoniou, F.
K. Diakonos, G. Mavromanolakis, arXiv:0912.4198 [nucl-ex].

\bibitem{Aggarwal2010}
M. M. Aggarwal \textit{et al.} (STAR Collaboration), Phys. Rev.
Lett. 105 (2010) 022302.

\bibitem{Aggarwal2010b}
M. M. Aggarwal \textit{et al.} (STAR Collaboration), arXiv:1007.2613
[nucl-ex].

\bibitem{Stephanov1998}
M.~A.~Stephanov, K.~Rajagopal, E.~V.~Shuryak, Phys. Rev. Lett. 81
(1998) 4816.

\bibitem{Stephanov2010}
M. A. Stephanov, Phys. Rev. D 81 (2010) 054012.

\bibitem{Singh2010}
A.~Singh, S.~Puri, H.~Mishra, arXiv:1004.3368 [hep-ph].

\bibitem{Schwinger1961}
J.~Schwinger, J. Math. Phys. 2 (1961) 407.

\bibitem{Keldysh1965}
L.~V.~Keldysh, Sov. Phys. JETP 20 (1965) 1018.

\bibitem{Chou1985}
K.~C.~Chou, Z.~B.~Su, B.~L.~Hao, L.~Yu, Phys. Rep. 118 (1985) 1.




\bibitem{Klevansky1997}
S.~P.~Klevansky, A.~Ogura, J.~H\"{u}fner, Annals Phys. 261 (1997)
37.





\bibitem{Zhou1980}
G.~Z.~Zhou, Z.~B.~Su, B.~L.~Hao, L.~Yu, Phys. Rev.  B 22 (1980)
3385.

\bibitem{Calzetta1988}
E.~Calzetta, B.~L.~Hu, Phys. Rev.  D 37 (1988) 2878.

\bibitem{Hall1975}
A.~Hall, J. Phys. A 8 (1975) 214.

\bibitem{Zhou1981}
G.~Z.~Zhou, Z.~B.~Su, Ch. 5 in ``Progress in Statistical Physics
(Chinese)'', eds. B.~L.~Hao \textit{et al.}, KEXUE (Science Press),
Beijing, p.268 (1981).

\bibitem{Zhou1982}
G.~Z.~Zhou, Z.~B.~Su, B.~L.~Hao, L.~Yu, Commun. in Theor. Phys.
(Beijing, China)  1 (1982) 307.

\bibitem{Nambu1961}
Y.~Nambu, G.~Jona-Lasinio, Phys. Rev.  122 (1961) 345; Phys. Rev.
124 (1961) 246.

\bibitem{Volkov1984}
M.~K.~Volkov, Ann. Phys. 157 (1984) 282.

\bibitem{Klevansky1992}
S.~P.~Klevansky, Rev. Mod. Phys. 64 (1992) 649.

\bibitem{Hatsuda1994}
T.~Hatsuda, T. Kunihiro, Phys. Lett. B 145 (1984) 7; Phys. Rep. 247
(1994) 221.

\bibitem{Alkofer1996}
R.~Alkofer, H. Reinhardt, H. Weigel, Phys. Rep. 265 (1996) 139.

\bibitem{Buballa2005}
M.~Buballa, Phys. Rep. 407 (2005) 205.

\bibitem{Zhou1982c}
G.~Z.~Zhou, Z.~B.~Su, B.~L.~Hao, L.~Yu, Commun. in Theor. Phys.
(Beijing, China) 1 (1982) 389.

\bibitem{Jackiw1974}
R.~Jackiw, Phys. Rev. D 9 (1974) 1686.

%\cite{Kunihiro:1995rb}
\bibitem{Kunihiro:1995rb}
  T.~Kunihiro,
  %``Chiral Symmetry And Scalar Meson In Hadron And Nuclear Physics,''
  Prog.  Theor.  Phys.  Suppl.  120 (1995) 75.
%  [arXiv:hep-ph/9502305].
  %%CITATION = PTPSA,120,75;%%

%\cite{Hatsuda:1985eb}
\bibitem{Hatsuda:1985eb}
  T.~Hatsuda, T.~Kunihiro,
  %``Fluctuation Effects In Hot Quark Matter: Precursors Of Chiral Transition At
  %Finite Temperature,''
  Phys. Rev. Lett.  55 (1985) 158.
  %%CITATION = PRLTA,55,158;%%




\end{thebibliography}



\end{document}